\documentclass[12pt]{iopart}
\usepackage{epsfig}
\begin{document}
%
%
\title[Heterostructures of strongly correlated materials]{Charge transfer 
in heterostructures of strongly correlated materials}
\author{I~Gonz\'alez$^{1,2}$, S~Okamoto$^{2}$, S~Yunoki$^{1,2}$, 
A~Moreo$^{1,2}$ and E~Dagotto$^{1,2}$}
\address{$^1$~Department of Physics and Astronomy, The University of Tennessee, 
Knoxville, TN 37996, USA}
\address{$^2$~Materials Science and Technology Division, Oak Ridge National 
Laboratory, Oak Ridge, TN 37831, USA}
%
%
\submitto{\JPCM}
%
%
\begin{abstract}
In this manuscript, recent theoretical investigations by the authors in
the area of oxide multilayers are briefly reviewed. The calculations 
were carried out using model Hamiltonians and a variety of non-perturbative
techniques. Moreover, new results are also included here. They correspond
to the generation of a metallic state by mixing insulators in a multilayer
geometry, using the Hubbard and Double Exchange models. For the latter,
the resulting metallic state is also ferromagnetic. This illustrates how
electron or hole doping via transfer of charge in multilayers can lead to
the study of phase diagrams of transition metal oxides in the clean limit.
Currently, these phase diagrams are much affected by the disordering
standard chemical doping procedure, which introduces quenched disorder in
the material.
\end{abstract}

\section{Introduction}

The study of heterostructures involving strongly correlated materials have
attracted considerable attention recently~\cite{ohtomo02}. One of the main
subjects of interest is the possibility of stabilizing new phases at the
interface between two transition metal oxides. In general, these two
materials will have different work functions creating a situation of
non-equilibrium. The electronic system reacts to the mismatch of the work
functions by generating an inhomogeneous charge distribution at the
interface resulting in some electronic charge being transferred between
the two materials. The electrostatic potential created by the
inhomogeneous charge distribution compensates the difference in the work
functions.

In principle, this physics appears to be quite similar to that found in
interfaces of semiconductors. However, strongly correlated materials have
complex phase diagrams with very different competing phases as the
electronic charge density, pressure, temperature, and external fields are
varied~\cite{dagotto}. From this perspective, interfaces of oxides have
considerable potential to create novel physics.

The goals of this paper are the following: (1) first, we will briefly
review previous theoretical work by the authors in the area of modeling
and computer simulations, addressing the effect of the charge transfer in
several on these oxide heterostructures. In particular, we will focus on
the possibility of electronic doping in these heterostructures, namely
reaching  electronic densities intermediate between those of insulators,
in a region of the material which is chemically homogeneous. Electronic
doping, as opposed to chemical doping, does not induce structural or
Coulombic defects in the material. Thus, effects of quenched disorder can
be studied, raising the possibility of reaching higher critical
temperatures in heterostructures than in chemically-doped bulk materials.
(2) The second goal of this paper is to present new results related with
the mixture (in multilayer geometries) of insulating antiferromagnets. It
is observed that this mixture can lead to a metal with very different
magnetic properties than the constituents. A simple model and calculation
illustrates the physics that induces this interesting behavior.

The outline of this paper is the following. In sections~\ref{sec:satoshi}
and~\ref{sec:oldResults}, we briefly review the charge transfer at
interfaces of Ti oxides and Cu/Mn oxides. In section~\ref{sec:DMRG}, new
results are presented. Here, we study the charge transfer that takes place
in a heterostructure formed by alternating layers of two insulators
described by the Hubbard and Double Exchange (DE) models. We analyze the
case when the layers are thin enough to allow the charge to be transferred
all throughout the heterostructure, leading to a metallic state.

Note that the brief nature of this manuscript does not allow us to fully
review the rapidly growing field of oxide interfaces. We recommend the
reader to consult the original publications by the authors, such 
as~\cite{okamoto04a,yun07}, for a broader view of this area of research.

\section{\label{sec:satoshi}
Electronic reconstruction at Mott-insulator/band-insulator
heterostructures}

In this section, we review the theoretical work that explained the
existence of a metallic phase at the interface of two Ti-oxide materials.
At present, it is widely recognized that interfaces between different
correlated electron systems can generate new electronic phases that are
different from the bulk. Let us discuss this rather general concept,
namely {\em electronic reconstruction}, by using a model heterostructure.
Specifically, we consider a [001]-type heterostructure in which a Mott
insulator and a band insulator with cubic perovskite structure ABO$_3$ are
grown along the $z$ direction. This corresponds to the LaTiO$_3$/SrTiO$_3$
heterostructures reported by Ohtomo~{\it et al.}~\cite{ohtomo02}. We
define a model heterostructure by placing +1 point charges at some of the
A sites. The charge +1 corresponds to the charge difference between a
rare-earth ion (charge +3) and an alkaline-metal ion (charge +2).
Electrons are assumed to move between nearest-neighbor B sites involving
transition-metal $d$-shells. They suffer an on-site Hubbard interaction
$U$, long-range repulsive Coulomb interactions with electrons on different
sites, and also attractive interactions with the +1 charged A ions. The
total electron number is determined by the neutrality condition; areal
densities of electrons and +1-charged A ions are equal. Thus,
A$^{+2}$BO$_3$ (no electrons) is a band insulator characterized by an
empty conduction band above the Fermi level. On the other hand, when the
on-site interaction is substantially strong, A$^{+3}$BO$_3$ (one electron
per site) becomes a Mott insulator characterized by Hubbard bands centered
at $\pm U/2$ separated by a Mott gap.

\begin{figure}[tbp]
\begin{center} 
\includegraphics[width=9cm,clip]{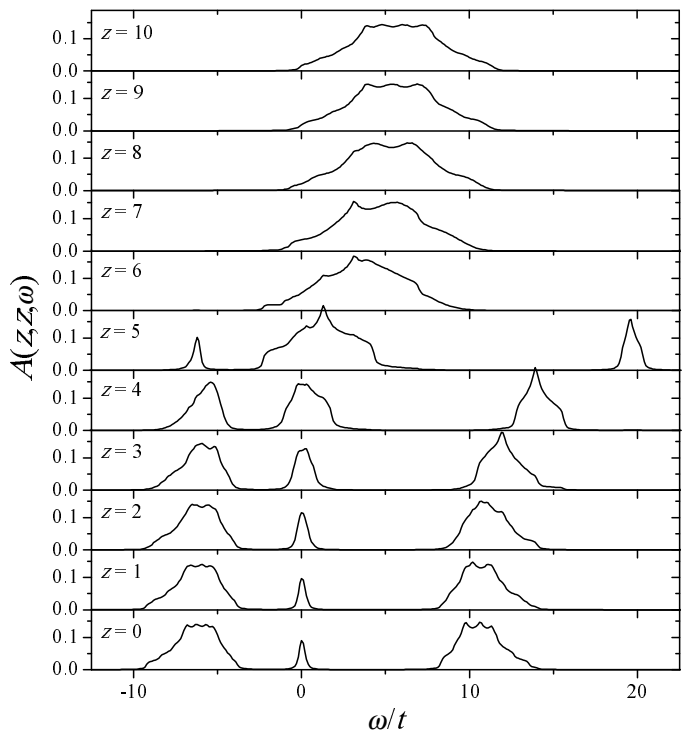}
\end{center} 
\caption{\label{fig:spectra} 
Spatially-resolved spectral function of a model
Mott-insulator/band-insulator heterostructures~\cite{okamoto04a}.
Parameters are chosen as $U=16t$ for the on-site Hubbard interactions (the
hopping integral is $t$). For the long-range Coulomb interactions, a
dielectric constant $\varepsilon =15$ and a lattice constant $a=4$~\AA
were used. The heterostructure is defined by +1 charges placed at the
A-site layers at $z=\pm 0.5,\pm 1.5,...\pm 4.5$ so the electronic (B)
sites are located at integer values of $z$. More details can be found
in~\cite{okamoto04a}.}
\end{figure}

Dealing simultaneously with strong correlations and spatial inhomogeneity
is a theoretical challenge. For this purpose, we generalized the dynamical
mean field theory (DMFT)~\cite{georges96} to a multilayer
geometry~\cite{okamoto04a,potthoff2003}, and solved the self-consistency
equations with a Hartree approximation for the long-range part of the
Coulomb interactions. Figure~\ref{fig:spectra} shows a numerical result
for the spatially-resolved spectral function of electrons, for a model
Mott-insulator/band-insulator heterostructure~\cite{okamoto04a}. Layers at
$|z| < 5$ and at $|z| > 5$ correspond to a Mott-insulating region and
band-insulating regions, respectively, and the layer at $|z| = 5$ is the
interface. At layers $|z| \gg 6$, the spectral function is essentially
identical to that of a bulk band-insulator. Approaching the
Mott-insulating region by reducing $z$, the spectral function evolves
fairly continuously. Eventually, the conduction band turns into a sharp
quasiparticle band at the Fermi level, dominating the spectral weight at
the interface layer $|z|=5$. The existence of finite spectral weight at
the Fermi level indicates the metallic property of the heterostructure.
Penetrating into the Mott-insulating region $|z|<5$, the quasiparticle
bands loses its weight exponentially. These behaviors contrast with a
simple band-bending picture for interfaces between two band insulators
with a finite band offset. 

The metallic behavior of such Mott-insulator/band-insulator heterostructures 
was actually reported experimentally in early work by Ohtomo {\it et
al}.~\cite{ohtomo02}. Furthermore, recent photoemission experiments on
LaTiO$_3$/SrTiO$_3$ superlattices have confirmed the appearance of a
quasiparticle band at the Fermi level, in agreement with the theoretical
prediction~\cite{takizawa06}. 

The essential physics controlling the properties described above is the
charge transfer between the two insulators. This creates intermediate
filling regions, in between the fillings of the two insulators, which are
responsible for the metallic behavior of the heterostructure. Therefore,
even if the on-site interaction is strong enough to produce a bulk Mott
insulator, metallic behavior survives at the interface. More realistic
model calculations including orbital degeneracy and electron-lattice
couplings further predicted interesting spin and orbital orderings that
are different from bulk materials~\cite{okamoto04a}. {\em Electronic
reconstruction}, which is the appearance of new electronic phases that are
different from the bulk electronic phases, at interfaces of correlated
electron systems is a quite general phenomenon. Interesting novel
electronic phases may result at  the interface of properly chosen oxides.
We will discuss such interface reconstructions in other models in the
following sections. 

\section{\label{sec:oldResults} Interfaces of manganites and cuprates}

The simple setup and results of the previous section show that the
transfer of charge between complex oxides should be a very general
phenomenon. In this section, recent theoretical efforts in this context
carried out by Yunoki {\it et al.}~\cite{yun07} are briefly reviewed (for
more details the reader should consult the original publication). The main
result of Ref.~\cite{yun07} was the prediction that a transfer of charge
could occur from a manganite to the {\it upper Hubbard band} of some
undoped cuprates (high-Tc parent insulators). Since electronic doping of
some Cu oxides has led to superconductivity, potentially the
manganite-cuprate multilayers discussed in~\cite{yun07} could also become
superconducting.

\begin{figure}[tbp]
\begin{center} 
\includegraphics[bb=47 348 568 640, clip=true, width=10cm]{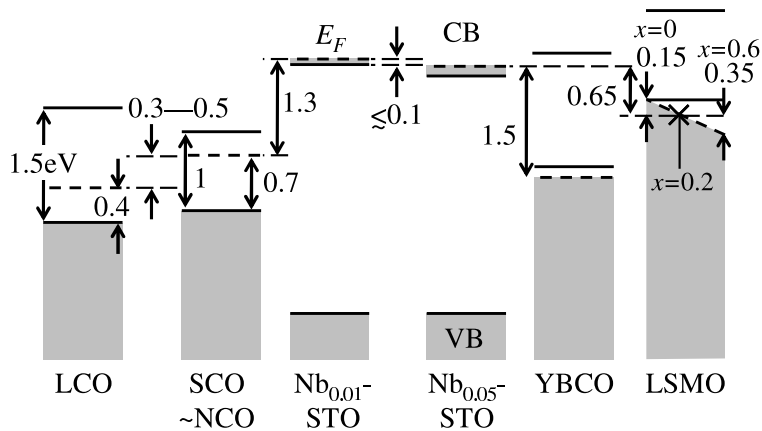}
\end{center} 
\caption{\label{fig:LCO_SCO_NbSTO_YBCO_LSMO} 
Schematic band diagrams of LCO, SCO(NCO), Nb$_{0.01}$-STO,
Nb$_{0.05}$-STO, YBCO, and LSMO based on diffusion voltage measurements
and photoemission spectroscopy. Tops of valence bands (VB) and bottoms of
conduction bands (CB) are indicated by solid lines, while chemical
potentials are indicated by dashed lines. This figure and caption are
reproduced from~\cite{yun07}, where more details can be found.
}
\end{figure}

One of the main results of~\cite{yun07} is the discussion of a band
alignment study of several oxides, which is illustrated in
Fig.~\ref{fig:LCO_SCO_NbSTO_YBCO_LSMO}. Using experimental information,
such as the work functions of oxides, the relative positions of the Mott
gaps and the chemical potentials were (crudely) predicted. This plot
suggests that the mixture of a manganite, such as La$_{1-x}$Sr$_x$MnO$_3$
(LSMO), and a doped superconducting cuprate, such as YBa$_2$Cu$_3$O$_y$
(YBCO), should lead to the transfer of charge from LSMO to YBCO. This is
in agreement with recent experimental results~\cite{keimer}, giving
confidence to the qualitative validity of the analysis. Moreover, new
predictions can be made. For instance, mixing LaMnO$_3$ (LMO) with 
Sm$_2$CuO$_4$ (SCO) should lead to the transfer of charge from LMO to the
upper band of SMO, and probably to an electron-doped superconductor, as
already mentioned. While more details about this particular case can be
found in~\cite{yun07}, here the main issue to remark is that by the
procedure outlined in this section and~\cite{yun07}, it is possible to
make qualitative predictions for the direction of charge transfer at
interfaces. This is of fundamental importance for the guidance of
experimental efforts, since the number of combinations of oxides is
enormous and theory must predict which of those combinations are
potentially the most attractive for the fabrication of superlattices.

\begin{figure}[hbt]
\begin{center}
\epsfig{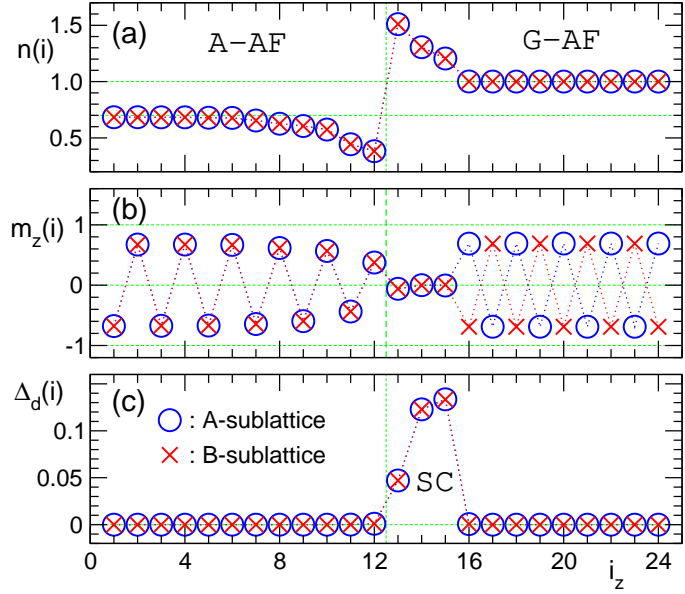}
\end{center}
\caption{\label{fig:a-af_sc} Transfer of charge from an A-type AF state
(as in some doped manganites~\cite{dagotto-CMR}) to an AF insulator (such
as undoped LCO, SCO, NCO, or YBCO), inducing  an electron-doped SC state
at the interface. The actual DE model parameters used here are $J_H=8t$,
$t_z=t$, $W_{\rm L}=14t$, and $n_+^{\rm L}=0.7$ for the left side of the
system, and $U=4t$, $V=-3t$, $t_z=0.1t$, $W_{\rm R}=0$, and $n_+^{\rm
R}=1.0$ for the right side of the system. $\alpha=1$ is set for the whole
system, with $L=16\times16\times24$ being the lattice studied. The
interface is located at $i_z=12.5$. The localized spins in the left side
of the systems are fixed to be in an A-type antiferromagnetic state, and
the temperature of the study was $T$=$t$/400. This figure and caption are
reproduced from~\cite{yun07}, where the reader can find more details,
including the model Hamiltonian used.
}
\end{figure}

The intuitive picture based on the work functions, and the possible
development of superconductivity in some cuprates via a proximity with the
manganites, was further substantiated in~\cite{yun07} by actual numerical
calculations. For example, Fig.~\ref{fig:a-af_sc} illustrates the results
of a simple mean-field approximation. The upper panel shows the electronic
density vs. position along the chain, for the case of an interface between
an A-type AF state (simulating an undoped manganite) and a G-type AF state
(simulating an undoped cuprate), after a Poisson equation iterative
procedure is carried out. Details can be found in~\cite{yun07}. The middle
panel illustrates the magnetic properties. The spin arrangement away from
the interface is either A or G, as expected by construction. However, near
the interface on the G-AF side (which simulates the undoped cuprate) an
accumulation of electronic charge takes place due to the different
chemical potentials of the two materials. This region is electron doped,
reducing drastically the spin G-type AF tendencies. Concomitant with this
reduction of antiferromagnetism, the lower panel shows the expected
development of superconductivity. Thus, the theoretical calculations are
in agreement with the simple picture based on the work functions. However,
note that more sophisticated calculations, beyond the mean-field
approximation, are needed to fully confirm these tendencies.

\begin{figure}[hbt]
\begin{center}
\epsfig{file=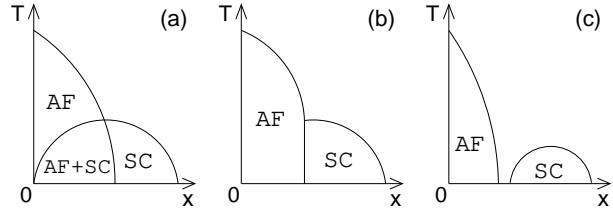,width=8cm}
\end{center} 
\caption{\label{fig:phase_SC}(a) and (b) are the possible phase diagrams
of the cuprates in the clean limit~\cite{alvarez}, which may be
experimentally realized at the interfaces discussed in \cite{yun07} and
briefly reviewed here. No distinction is made between hole or electron
doping, $x$ represents both. (c) is the well-known phase diagram of
chemically doped La$_{2-x}$Sr$_x$CuO$_4$ (LSCO), which can also be
obtained from (a) and (b) by adding quenched disorder. According
to~\cite{alvarez}, the glassy state between AF and SC phases is caused by
quenched disorder, and it contains superconducting and antiferromagnetic
clusters.}
\end{figure}

The possibility of generating an electron doped superconductor via charge
transfer from other oxides may help in unveiling the true phase diagram of the
high temperature superconductors. According to phenomenological calculations
by Alvarez {\it et al.}~\cite{alvarez}, in the absence of quenched disorder
(clean limit) the phase diagram of cuprates should have either a
first-order transition separating the competing states or a region of
local coexistence of both orders (actually, stripes are a third more
exotic possibility, see~\cite{alvarez}). None corresponds to the true 
(experimental) phase diagram of LSCO. The clean-limit proposed and actual
experimentally observed phase diagrams are in Fig.~\ref{fig:phase_SC}. The
key idea of~\cite{alvarez} was the observation that the LSCO true phase
diagram can be obtained from the clean limit phase diagram by merely
adding quenched disorder. This opens a window in the phase diagram between
the competing phases and induces an intermediate glassy state. If this
observation is correct, then quenched disorder fundamentally affects the
cuprate's phase diagrams. Then, the possibility of doping Cu-oxide parent
insulators via charge transfer in multilayers becomes a possible path to
reveal the real clean-limit phase diagrams of the cuprates, which at
present may be much distorted by the chemical doping procedure.

\section{\label{sec:DMRG} Superlattices of insulating materials can lead
to a metal}

Besides containing a brief review of recent work on interfaces of
correlated electrons, this manuscript also includes new results that are
described in this section. The goal here is to illustrate, with a simple
example, how an array (multilayer) of thin layers (thin meaning just 
a few lattice spacings in width) can have properties drastically different
from those of the bulk constituents. In particular, the cases of the
Hubbard and DE models will be investigated. In both examples, the
``building blocks'', namely the isolated layers, are insulating. However,
it will be shown that the ensemble becomes {\it metallic} due to transfer
of charge.

\subsection{Superlattices in the Hubbard model}

In this subsection, the results for a superlattice described by the
well-known Hubbard model are presented. The heterostructure studied here
is formed by alternating layers of two different materials, labeled A and
B, which are chosen to be insulating and chemically homogeneous. In the
simple Hubbard electronic Hamiltonian, each material can be parametrized
by selecting its electronic density, which in the bulk locally matches the
charge of the positive ions for an homogeneous system. Let us assume that
only one band is relevant to determine the properties of the material
under study: then the system can be described by the (single-orbital)
Hubbard model defined by 

\begin{equation}\label{eq:hamiltonian}
\hat H = -t\sum_{\langle i,\,j\rangle}c_{i\sigma}^\dagger c_{j\sigma}
+U\sum_{i} n_{i\uparrow}n_{i\downarrow} 
+\sum_{i}\left(V_{i}-\mu+W_{\rm A/B} \right)n_{i}, 
\end{equation}   
where $c^{\dagger}_{i\sigma}$ is the creation operator for an electron with
spin projection $\sigma=\uparrow,\downarrow$ at site $i$, $t$ is the
hopping integral between neighboring sites, and $\langle i, j\rangle$
denotes nearest neighbors; $n_{i}=n_{i\uparrow}+n_{i\downarrow}$
gives the number of electrons at site $i$, $U$ is the on-site Coulomb
repulsion, and $V_{i}$ is the electronic potential (discussed in more detail
later) that will take into account effects related to the charge
redistribution. $\mu$ is the chemical potential, while $W_{\rm A}$ and $W_{\rm
B}$ are site potentials. Finally, to simplify the numerical task, 
without altering the qualitative aspects of the conclusions, a one-dimensional
arrangement will be studied. The simplicity of the results described below
lead us to believe that the conclusions are valid in higher dimensions as well.

\begin{figure}
\begin{center}
\epsfig{file=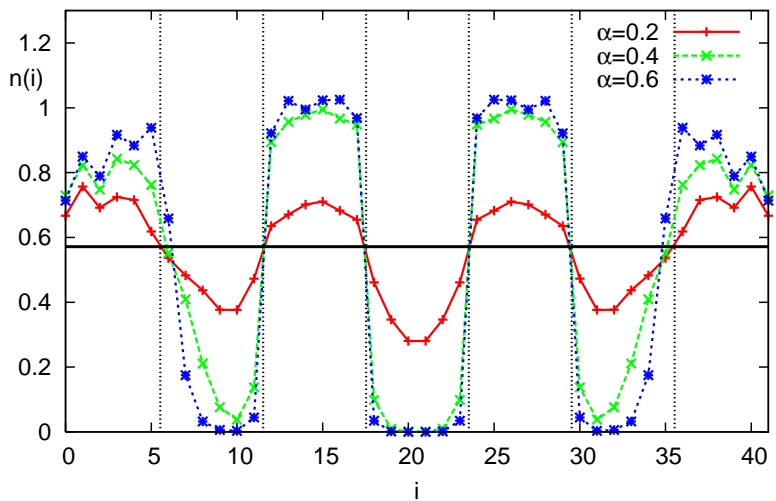, clip=true, width=8cm,angle=-90}
\end{center}
\caption{\label{fig:one_term} Electronic charge density $\langle
n_i\rangle$ along the one-dimensional structure shown (that simulates a
multilayer), for several values of the reciprocal of the dielectric
constant $\alpha$. The parameters of the
Hamiltonian~(\ref{eq:hamiltonian}) are $U=4t$ and $W_{A}=W_{B}$. The
heterostructure is formed by alternating 6-site layers with ionic
backgrounds $n^{+}_{A}=1,\, n^{+}_{B}=0$. The average electronic density
is $n=0.5714$. For small enough $\alpha$, the charge is fairly
homogeneously distributed and the ensemble is expected to be metallic.
} 
\end{figure}

Note that $V_{i}$ has contributions coming from both the ionic and
electronic charges. While the ionic part can be easily calculated, 
the dependence on the electronic densities makes $V_{i}$ a complicated 
operator. In the continuum limit, $V_{i}$ should be determined by the Poisson
equation. Therefore, we can use an iterative procedure to calculate the
ground state properties of the
Hamiltonian~(\ref{eq:hamiltonian})~\cite{oka}. For a given iteration
{\it it}, we assume that we have a guess for the electronic charge
distribution $n_{i}^{it}$ that is used to calculate $V_{i}^{it}$ by
solving the discretized Poisson equation:

\begin{equation}
V_{i+1}-V_{i-1}-2V_{i}=\alpha\left(n_{i}-n^{+}_{i}\right),
\end{equation}
where $\alpha=e/a\varepsilon$, $\varepsilon$ is the dielectric constant,
$e$ is the charge of the electron $e$, and $a$ is the lattice constant.
The right-hand-side of the equation is a lattice discretized
version of the second derivative operator. The ground state of the
Hamiltonian~(\ref{eq:hamiltonian}), $\psi^{it}$, is calculated via the
DMRG algorithm~\cite{white} using $V^{it}_i$. The value for 
the next iteration is calculated as $n_{i}^{it+1}=\sum^{N}_{j=0}\beta_{j}\langle \psi^{it-j} | n_{i} \psi^{it-j}\rangle$, where $\beta_{j}\in(0,1]$ and
$\sum_{j}\beta_{j}=1$. The procedure is repeated until the set $V^{it}_i$
converges.

To calculate the ground state we typically keep 200 states per 
DMRG block. We perform enough number of DMRG sweeps between two consecutive 
solutions of the Poisson
equation to verify that each  $|\psi^{it}\rangle$ is converged for the
$V^{it}_{i}$ used. In practice, typically $\beta_{0}\approx0.9$ and
$N\approx2$. We have found empirically that the iterative procedure is
particularly difficult to converge for $\alpha>0.5$, if $\beta<0.9$.

\begin{figure}
\begin{center}
\epsfig{file=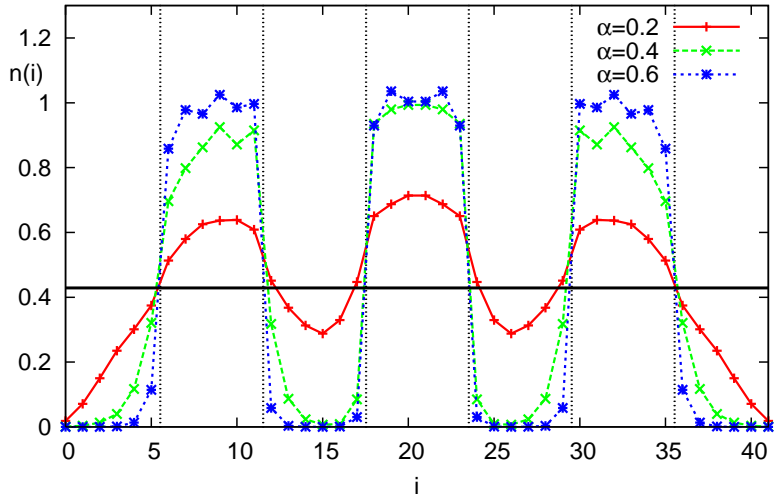,width=8cm,angle=-90}
\end{center}
\caption{\label{fig:zero_term} Electronic charge density $\langle
n_i\rangle$ along the multilayer structure shown, for several values of
the reciprocal of the dielectric constant $\alpha$. The values of the
parameters are the same as in Fig.~\ref{fig:one_term}, with the exception
of the ionic background which were inverted and are now $n^{+}_{A}$=0 and
$n^{+}_{B}$=1. The average electronic density is $n=0.4285$. 
}
\end{figure}

Typical results are shown in  Fig.~\ref{fig:one_term} where the charge
density profile along the chain, parametric with the reciprocal value of
the dielectric constant $\alpha=1/\varepsilon$, is given for the case of
two insulating materials with $n^{+}_{A}$=1 and  $n^{+}_{B}$=0, and in
the realistic limit where $U$ is much larger than $t$ (namely, when a 
robust Hubbard gap develops at half-filling; $U$ is an atomic coupling
in the range of a few eVs, while $t$ is just a fraction of eV). From the
figure, it can be easily observed that the long-range Coulomb interaction,
considered via the Poisson equation, alters qualitatively the charge
profile from a highly inhomogeneous insulating state for large $\alpha$
(with electrons following the positive charge density of the bulk
materials) to a nearly {\it homogeneous metallic state} for small
$\alpha$. This last observation is the main result of this section: the
fact that the multilayer system is made out of thin layers allows the
electronic charge to be redistributed all along the heterostructure. From
previous investigations of the Hubbard model, it is known that this model
with electronic density $\sim$0.5 is metallic. Thus, by mixing insulators,
a metal emerges in the multilayer geometry for sufficiently thin layers, a
remarkable result.

For completeness, in Fig.~\ref{fig:zero_term} we also show results for a
similar electronic charge density profile but for  the case when the ionic
background charges are $n^{+}_{A}$=0 and $n^{+}_{B}$=1. As before, the
charge is effectively redistributed all along the chain~\cite{sarma}.

\subsection{\label{sec:MFT} Superlattices in the Double Exchange model}

\begin{figure}
\begin{center}
\epsfig{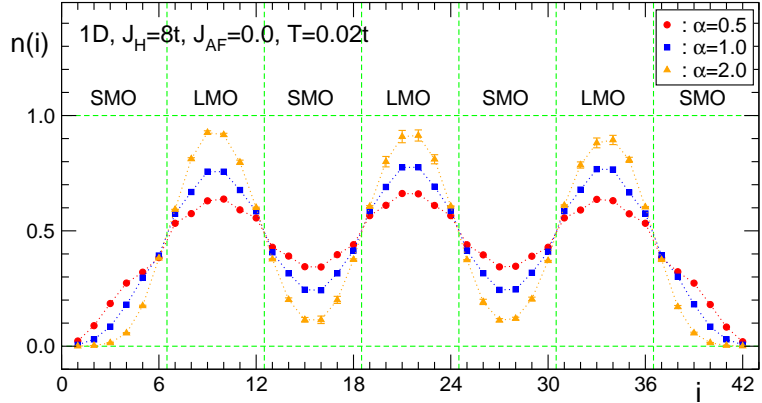}
\end{center}
\caption{\label{fig:Seiji_2} Electronic charge vs. position along
a chain of 42 sites, using the DE model and parameters and temperature
indicated. Clearly the charge spreads as $\alpha$ is reduced.}
\end{figure}

The previous Hubbard model results could be used for interfaces involving
a Mott insulator and a band insulator. However, if one of the building
blocks is a manganite, then the DE model should be employed. This DE model
takes into account that the electrons that form the bands close to the
Fermi level are in partially filled $d$-orbitals. The degeneracy of the
$d$-orbitals is usually removed by the crystal field created by the
underlying crystalline structure. This results in the appearance of a gap
between the different representations of the $d$-orbitals in the symmetry
group of the ionic lattice, for example between the $t_{\rm 2g}$ and
$e_{\rm g}$ orbitals in the case of the perovskite tetragonal structure.
The filled sub-band can be described by a localized spin. In the case of
interest here (LaMnO$_3$ (LMO) and SrMnO$_3$ (SMO)), the $e_{\rm g}$
sub-band is higher in energy than the $t_{\rm 2g}$ sub-band, which is
filled and it is represented by a localized $3/2$-spin (usually assumed
classical for simplicity). Then, a manganite system can be described by
the DE Hamiltonian~\cite{dagotto-CMR}:

\begin{eqnarray}\label{eq:DE_hamiltonian}
\hat H &=& -t \sum_{\langle i,\,j\rangle}\sum_{\sigma}
c_{i\sigma}^\dagger c_{j\sigma}
+\sum_{i}\left(V_{i}-\mu+W_{\rm L/R} \right)n_{i} \nonumber\\
& & -J_{\rm H}\sum_{i}\sum_{\alpha,\beta}
c_{{i}\alpha}^{\dag}\left(\vec{\sigma}\right)_{\alpha\beta}
c_{i\beta}\cdot {\vec{S}}_{i}
+\frac{J_{\rm AF}}{2}\sum_{\langle i,\,j\rangle}\vec{S}_{i}\cdot\vec{S}_{j},
\end{eqnarray}   
where now $c^{\dagger}_{i,\sigma}$ creates an $e_{\rm g}$-electron at site
$i$ with spin projection $\sigma=\uparrow,\downarrow$. For simplicity,
here only one band in the $e_{\rm g}$ manifold is used, a widely used
approximation. $t$ is the hopping integral between neighboring sites for
the electrons in this $e_{\rm g}$ sub-band; $J_{\rm H}$ is the Hund
coupling between the electrons in the $e_{\rm g}$ sub-band; and $J_{\rm
AF}$ is the antiferromagnetic exchange interaction between neighboring
localized spins, which takes into account the virtual hoppings in the
$t_{\rm 2g}$ sub-band. $\vec{\sigma}=(\sigma_x,\sigma_y,\sigma_z)$ are
Pauli matrices, and ${\vec{S}}_{i}$ is a classical localized spin at site
{i} ($|{\vec{S}}_{i}|=1$) representing the $t_{\rm 2g}$ spins. More
details about this model and the physics of manganites in general can be
found in~\cite{dagotto-CMR}.

The considerations related to the long-range portion of the electrostatic
potential used in the previous subsection remain valid here as well, and a
similar iterative procedure is also used. The difference is that the
ground state of the Hamiltonian~(\ref{eq:DE_hamiltonian}) is solved by
the numerical Monte Carlo method, instead of the DMRG technique. The
Hamiltonian is separated in spin and electronic components. The electronic
portion is treated exactly via library subroutines. The localized spins
are treated in the classical approximation using a Monte Carlo algorithm~\cite{dagotto-CMR}.

Regarding the distribution of electronic charge, the results are
qualitatively similar to those in the case of the Hubbard model. A typical
case is shown in Fig.~\ref{fig:Seiji_2} at low temperature. Once again, as
$\alpha$ is reduced the charge spreads, and its value is far from the
nominal 1 and 0 of the building blocks. Clarifications are here in order:
(1) The coupling $J_{\rm H}$=8$t$ is realistic, as widely discussed in
previous literature ~\cite{dagotto-CMR}. In fact, $J_{\rm H}$ is estimated
to be a few eV's, while the hopping $t$ is always just a fraction of eV.
Note that $J_{\rm H}$ is a local on-site ferromagnetic interaction, and it is
not the actual parameter directly regulating the strength of the critical temperature.
The latter arises from an effective double-exchange coupling between nearest
neighbors spins. (2) The limit $J_{\rm AF}$=0 is chosen for simplicity.
It is already well-known~\cite{dagotto-CMR} that at $J_{\rm AF}$=0, and with one
$e_{\rm g}$-electron per site in the one-orbital model, the ground state
is antiferromagnetic. Then, LMO is properly described.
Regarding SMO, a better description would have needed $J_{\rm AF}$$\neq$0,
to represent the antiferromagnetism present in the limit of zero $e_{\rm g}$
carriers. However, this
approximation does not at all affect the conclusions (see below) of our effort.

\begin{figure}
\begin{center}
\epsfig{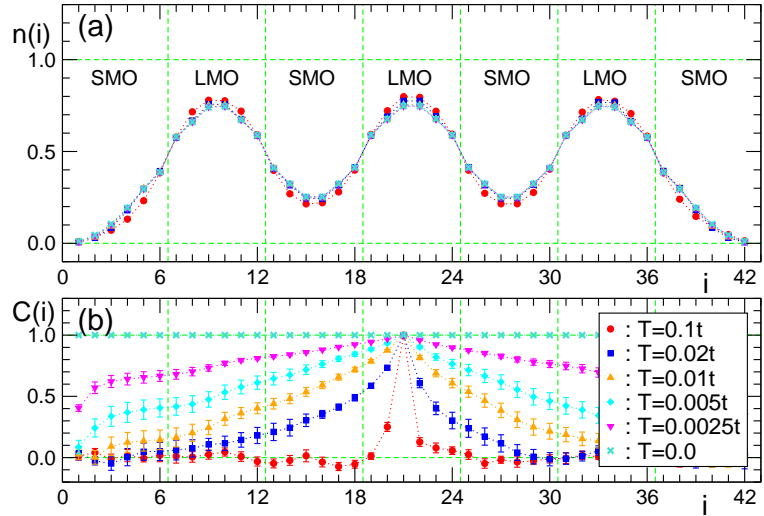}
\end{center}
\caption{\label{fig:Seiji_1} Ferromagnetism in multilayers of
antiferromagnetic manganites. Upper panel shows the charge distribution
for $\alpha$=1.0, which only has a weak temperature dependence. The lower
panel, on the contrary, shows the clear development of ferromagnetism upon
cooling. Shown are the spin-spin correlation from the center (site 21) to
the rest of the sites, $C_{i}=\langle\vec{S_{i}}\cdot\vec{S_{21}}\rangle$.
All results were obtained with the Monte Carlo technique, with the
exception of $T=0$ which was found using a minimization procedure for the
classical spins. The model was the DE, with the same parameters as in
Fig.~\ref{fig:Seiji_2}.
}
\end{figure}

For the case of the DE model, the magnetic properties are more interesting
than for the Hubbard model. The reason is that together with the metallicity
induced by charge transfer in a multilayer structure, the magnetic properties
change from antiferromagnetic (for the bulk components LMO and SMO) to
ferromagnetic (in the multilayer). This is illustrated in 
Fig.~\ref{fig:Seiji_1}, for $\alpha$=1. The upper panel shows the charge
density, which is not changing much in the range of temperatures
investigated. However, the lower panel, with the spin-spin correlations,
indicates clearly the development of {\it ferromagnetism} upon cooling.
The reason is simple. The electronic density is no longer 1 or 0, as in
the LMO and SMO limiting cases, but at every site this density becomes an
intermediate number between 1 and 0. This charge doping leads to
ferromagnetic tendencies, since the well-known DE mechanism becomes active
upon doping~\cite{dagotto-CMR}. In fact, it is known from previous
investigations that the tendencies toward ferromagnetism are the strongest
at electronic density 0.5, and they survive in a wide range of dopings.
From this perspective, the results are easy to understand: (1) the
long-range Coulomb interaction spreads the charge, effectively doping LMO
with holes and SMO with electrons; (2) in hole or electron doped manganite
antiferromagnets, the DE mechanism leads to ferromagnetism. However, there
is an important difference between multilayers and bulk compounds: the
chemical doping procedure usually employed to dope oxides, for instance
substituting La by Ca, is now replaced by a mere spreading of the charge
in the vicinity of the interface. Then, the influence of quenched disorder
is reduced by this procedure, as already discussed in the case of
superconductors in previous sections.

\begin{figure}
\begin{center}
\epsfig{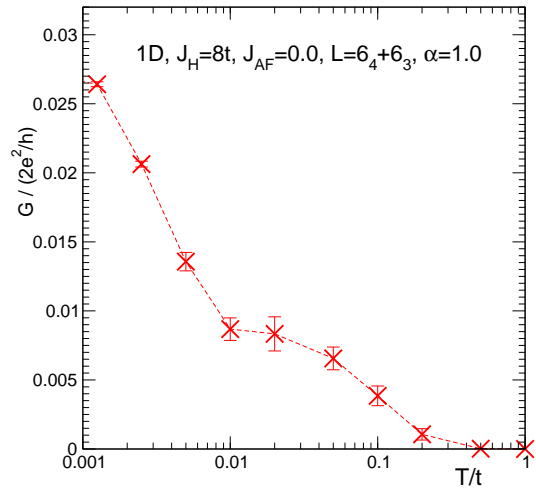}
\end{center}
\caption{\label{fig:Seiji_3} Conductance corresponding to the
one-dimensional structure used in Fig.~\ref{fig:Seiji_1}. The increase of
the conductance with reducing temperature indicates metallic behavior.
}
\end{figure}

The ferromagnetism goes together with metallicity, as shown in
Fig.~\ref{fig:Seiji_3} where the conductance is given as a function of
temperature. Similarly as it occurs in bulk DE models, the appearance of
ferromagnetism also leads to a substantial increase in the conductance.
The conductance increase with reducing temperature signals the existence
of a metal in the multilayer system that is made out of antiferromagnetic
insulators, which is a conceptually interesting result. Novel properties
arise in the multilayer structure as a whole that are not present in the
individual materials that form the multilayer.

\section{Conclusions}

In this manuscript, recent investigations by the authors in the area
of interfaces of complex oxides, using modeling techniques and numerical
simulations, were briefly reviewed. In addition, new results corresponding
to multilayers of insulating antiferromagnets (in a 1D arrangement for
simplicity) were also presented. The spreading of charge between the two
materials involved in the  multilayer is sufficiently strong to generate
a metallic state, which in the case of the manganites is ferromagnetic due
to the DE effect. These simple examples illustrate the potential of
working with oxide multilayers: they provide us with a novel procedure to
tune properties of materials by adjusting the width and the nature of the
components themselves. The number of combinations is huge and this field
of research is in its early stages. The experimental effort clearly
needs theoretical guidance to establish which are the most interesting
combinations of oxides to investigate.
This new ``playground'' for correlated electrons surely will provide
several surprises in the near future, which not only may influence on
fundamental research in complex oxides by generating new interfacial
phases, but may also be of potential relevance in devices in the growing
field of oxide electronics.

%
%
\ack
This work is supported in part by the NSF grant DMR-0706020 and the
Division of Materials Science and Engineering, U.S. DOE, under contract
with UT-Battelle, LLC.
%
%

\end{document}